\documentclass[a4paper]{jpconf}
\usepackage{graphicx}
\usepackage{epsfig}
\begin{document}
\title{From Little Bangs to the Big Bang}
\author{John Ellis}
\address{Theory Division, Physics Department, CERN, CH-1211 Geneva 23, Switzerland}
\ead{john.ellis@cern.ch ~~~~~ CERN-PH-TH/2005-070 ~~~~~ astro-ph/0504501}
\begin{abstract}

The `Little Bangs' made in particle collider experiments reproduce the
conditions in the Big Bang when the age of the Universe was a fraction of
a second. It is thought that matter was generated, the structures in the
Universe were formed and cold dark matter froze out during this very early
epoch when the equation of state of the Universe was dominated by the
quark-gluon plasma (QGP). Future Little Bangs may reveal the mechanism of
matter generation and the nature of cold dark matter. Knowledge of the QGP 
will be an essential ingredient in quantitative understanding of the very 
early Universe.

\end{abstract}.

\section{The Universe is Expanding}

The expansion of the Universe was first established by Hubble's discovery
that distant galaxies are receding from us, with redshifts proportional to
their relative distances from us. Extrapolating the present expansion
backwards, there is good evidence that the Universe was once 3000 times
smaller and hotter than today, provided by the cosmic microwave background
(CMB) radiation. This has a thermal distribution and is very isotropic,
and is thought to have been released when electrons combined with ions
from the primordial electromagnetic plasma to form atoms. The observed
small dipole anisotropy is due to the Earth's motion relative to this
cosmic microwave background, and the very small anisotropies found by the
COBE satellite are thought to have led to the formation of structures in
the Universe, as discussed later~\cite{Oz}.

Extrapolating further back in time, there is good evidence that the
Universe was once a billion times smaller and hotter than today, provided
by the abundances of light elements cooked in the Big Bang~\cite{cfos}.  
The Universe contains about 24 $\%$ by mass of $^4$He, and somewhat less
Deuterium, $^3$He and $^7$Li. These could only have been cooked by nuclear
reactions in the very early Universe, when it was a billion times smaller
and hotter than today. The detailed light-element abundances depend on the
amount of matter in the Universe, and comparison between observations and
calculations suggests that there is not enough matter to stop the present
expansion, or even to explain the amount of matter in the galaxies and
their clusters. The calculations of the light-element abundances also
depend on the number of particle types, and in particular on the number of
different neutrino types. This is now known from particle collider
experiments to be three~\cite{LEP}, with a corresponding number of charged
leptons and quark pairs.

\section{The Very Early Universe and the Quark-Gluon Plasma}

When the Universe was very young: $t \to 0$, also the scale factor $a$
characterizing its size would have been very small: $a \to 0$, and the
temperature $T$ would have been very large, with characteristic
relativistic particle energies $E \sim T$. In normal adiabatic expansion,
$T \sim 1/a$, and, while the energy density of the Universe was dominated
by relativistic matter, $t \sim 1/T^2$. The following are some rough
orders of magnitude: when the Universe had an age $t \sim 1$ second, the
temperature was $T \sim 10,000,000,000$ degrees, and characteristic
thermal energies were $E \sim 1$ MeV, comparable with the mass of the
electron. It is clear that one needs particle physics to describe the
earlier history of the Universe~\cite{Oz}.

\begin{figure}[htb]
\centerline{\epsfxsize = 0.6\textwidth \epsffile{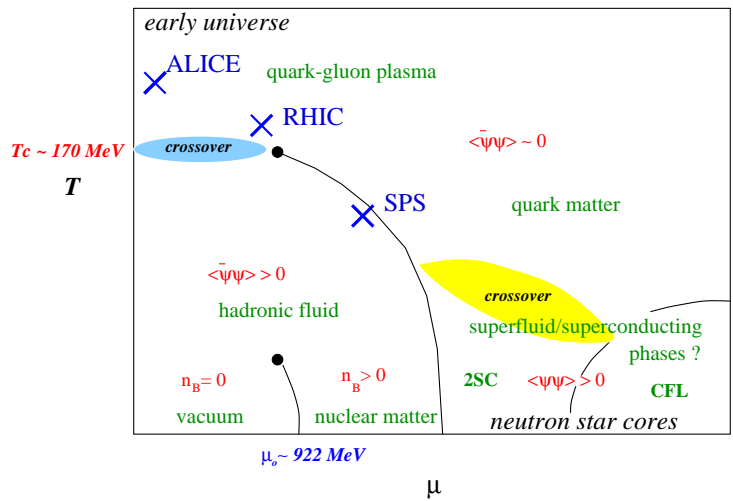}}
\caption{\it  
The phase diagram of hot and dense QCD for different values of the baryon 
chemical potential $\mu$ and temperature $T$~\protect\cite{Karsch}, 
illustrating 
the physics reaches of SPS, RHIC and the ALICE experiment at the 
LHC~\cite{ALICE}.}
\vspace*{0.5cm}
\label{fig:qcd}
\end{figure}

The very early Universe was presumably filled with primordial quark-gluon
plasma (QGP). When the Universe was a few microseconds old, it is thought
to have exited from this QGP phase, with the available quarks and gluons
combining to make mesons and baryons. The primordial QGP would have had a
very low baryon chemical potential $\mu$. Experiments with RHIC reproduce
cosmological conditions more closely than did previous SPS experiments, as
seen in Fig.~\ref{fig:qcd}, and the LHC will provide~\cite{ALICE} an even 
closer approximation to the primordial QGP. I shall not discuss here the
prospects for discovering quark matter inside dense astrophysical objects
such as neutron stars, which would have a much larger baryon chemical
potential.

\begin{figure}[htb]
\centerline{\epsfxsize = 0.5\textwidth \epsffile{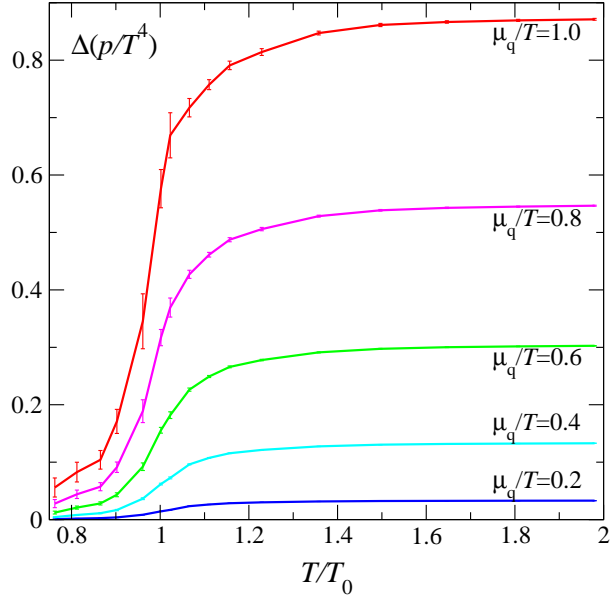}}
\caption{\it
The growth of the QCD pressure with temperature, for different values of 
the baryon chemical potential $\mu$~\protect\cite{Karsch}. The rise is 
quite smooth, indication that there is not a strong first-order phase 
transition, and probably no dramatic consequences in the early Universe.} 
\vspace*{0.5cm}
\label{fig:ledens}
\end{figure}

To what extent can information about the early Universe cast light on the
quark-hadron phase transition? The latest lattice simulations of QCD with
two light flavours $u, d$ and one moderately heavy flavour $s$ suggest
that there was no strong first-order transition. Instead, there was
probably a cross-over between the quark and hadron phases, see, for 
example, Fig.~\ref{fig:ledens}~\cite{Karsch}, during which
the smooth expansion of the Universe is unlikely to have been modified
substantially. Specifically, it is not thought that this transition would
have induced inhomogeneities large enough to have detectable consequences
today.

\section{Open Cosmological Questions}

The Standard Model of cosmology leaves many important questions
unanswered. {\it Why is the Universe so big and old?} Measurements by the
WMAP satellite, in particular, indicate that its age is about
14,000,000,000 years~\cite{Spergel}. {\it Why is its geometry nearly
Euclidean?} Recent data indicate that it is almost flat, close to the
borderline for eternal expansion. {\it Where did the matter come from?}
The cosmological nucleosynthesis scenario indicates that there is
approximately one proton in the Universe today for every 1,000,000,000
photons, and no detectable amount of antimatter. {\it How did cosmological
structures form?} If they did indeed form from the ripples observed in the
CMB, how did these originate? {\it What is the nature of the invisible
dark matter thought to fill the Universe?} Its presence is thought to have
been essential for the amplification of the primordial perturbations in
the CMB.

It is clear that one needs particle physics to answer these questions, and
that their solutions would have operated in a Universe filled with QGP.

\section{A Strange Recipe for a Universe}

\begin{figure}[htb]
\centerline{\epsfxsize = 0.8\textwidth \epsffile{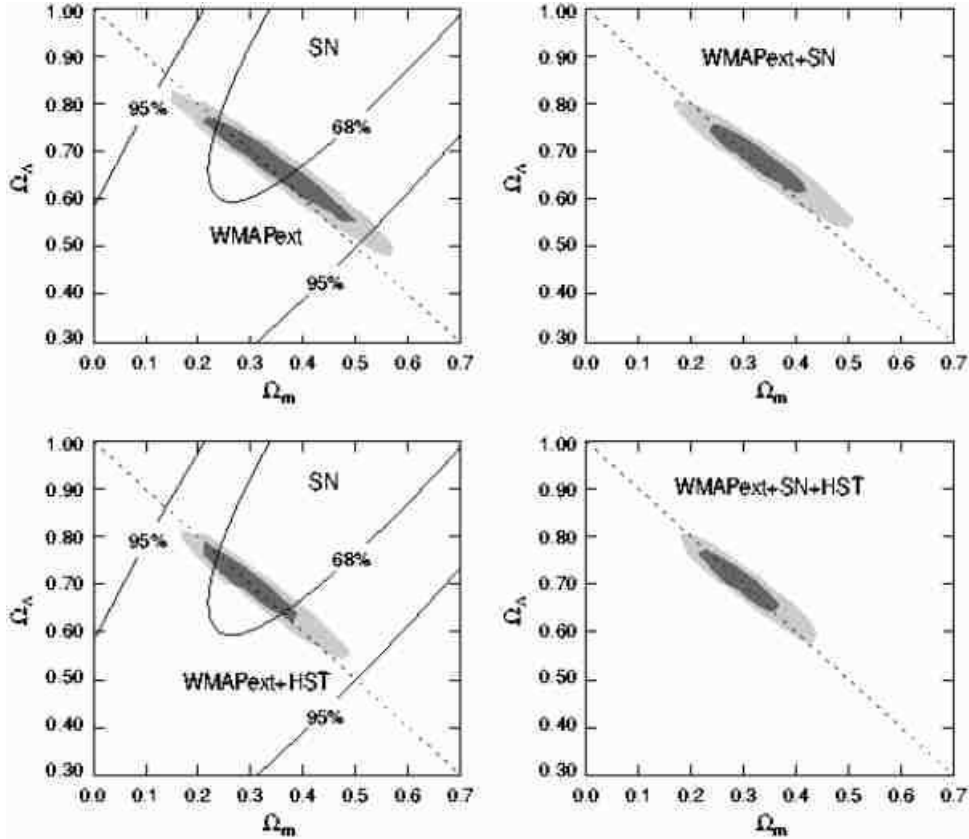}}
\caption{\it  
The density of matter $\Omega_m$ and dark energy $\Omega_\Lambda$
inferred from WMAP and other CMB data (WMAPext), and from combining them
with supernova and Hubble Space Telescope data~\protect\cite{Spergel}.}
\vspace*{0.5cm}
\label{fig:RS6}
\end{figure}

According to the `Concordance Model' suggested by a multitude of
astrophysical and cosmological observations, the total density of the
Universe is very close to the critical value: $\Omega_{Tot} = 1.02 \pm
0.02$, as illustrated in Fig.~\ref{fig:RS6}~\cite{Spergel}. The theory of
cosmological inflation suggests that the density should be
indistinguishable from the critical value, and this is supported by
measurements of the CMB. On the other hand, the baryon density is small,
as inferred not only from Big-Bang nucleosynthesis but also and
independently from the CMB: $\Omega_{Baryons} \sim \rm{few} \%$. The CMB
information on these two quantities comes from observations of peaks in
the fluctuation spectrum in specific partial waves corresponding to
certain angular scales: the position of the first peak is sensitive to
$\Omega_{Tot}$, and the relative heights of subsequent peaks are sensitive
to $\Omega_{Baryons}$. The fraction $\Omega_m$ of the critical density
provided by all forms of matter is not very well constrained by the CMB
data alone, but is quite tightly constrained by combining them with
observations of high-redshift supernovae~\cite{SN} and/or large-scale
structures~\cite{LSS}, each of which favours $\Omega_{Matter} \sim 0.3$,
as also seen in Fig.~\ref{fig:RS6}.

\begin{figure}
\begin{center}
 \includegraphics[height=.29\textheight, angle=0]{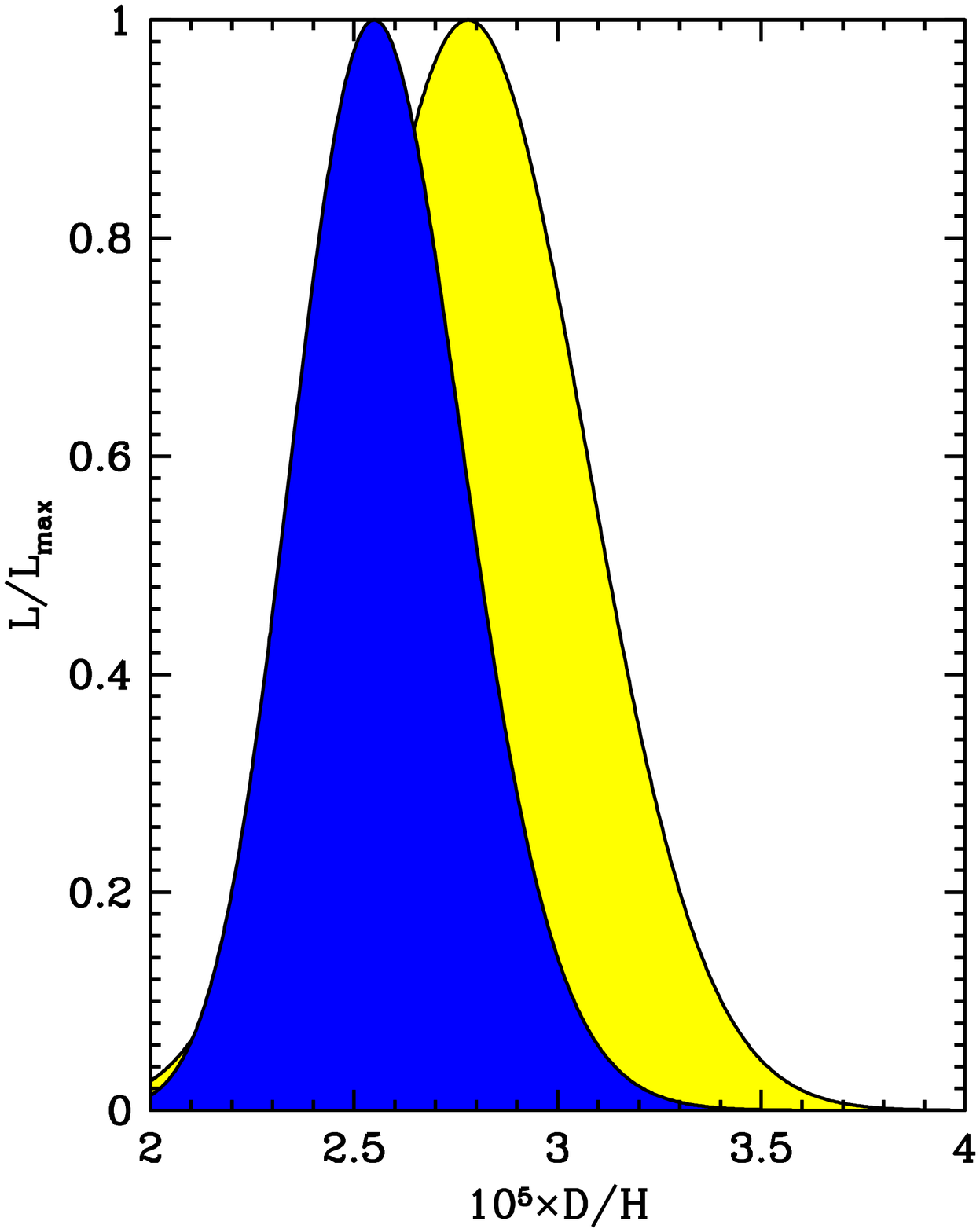}
  \includegraphics[height=.29\textheight, angle=0]{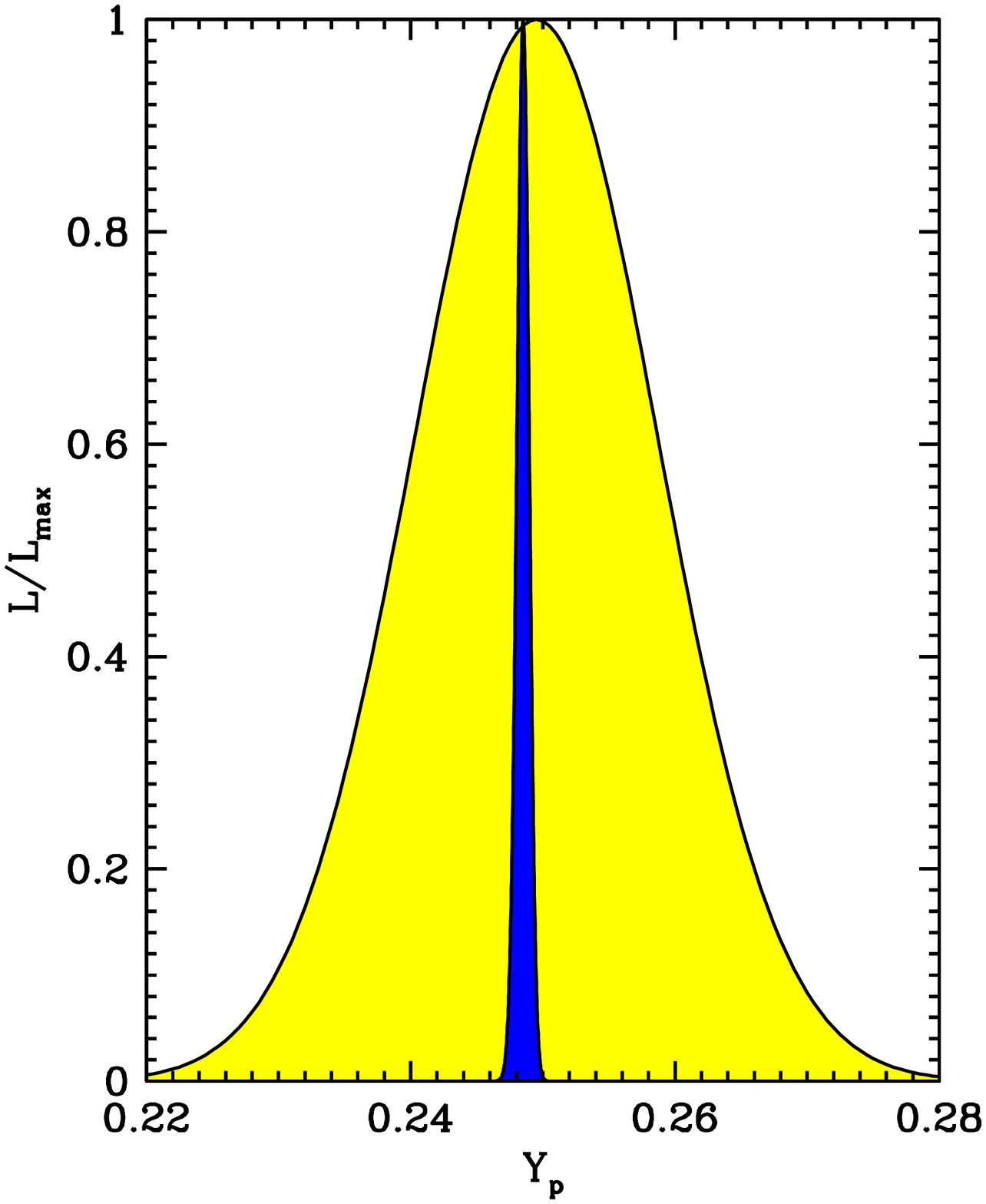}
   \includegraphics[height=.29\textheight, angle=0]{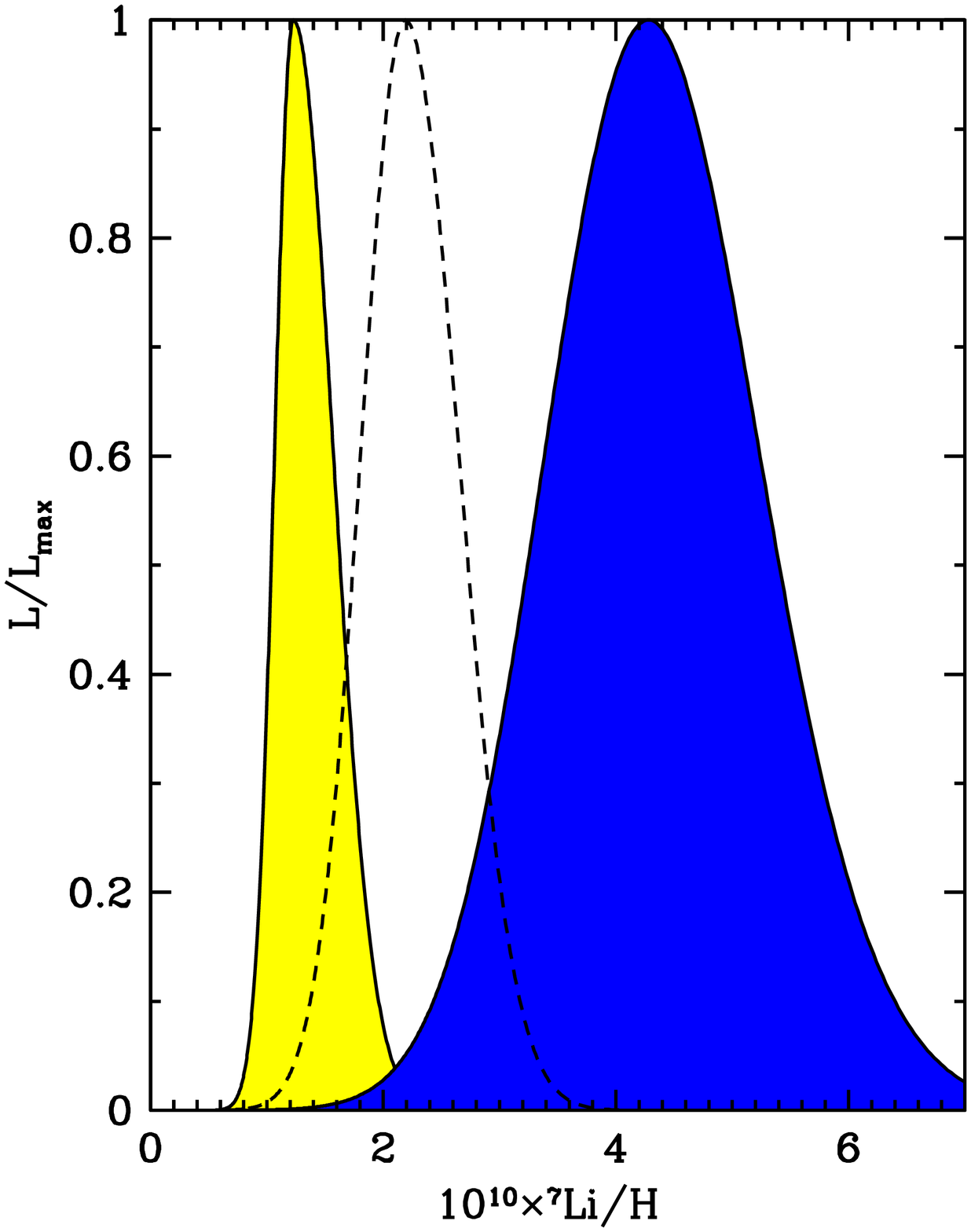}
\end{center}
\caption{\it
Primordial light element abundances as predicted by BBN (light) and
WMAP (dark shaded regions)\protect\cite{cfos}, for
(a) D/H, (b) the $^4$He abundance $Y_p$ and (c)
$^7$Li/H~\protect\cite{cfos}.
}
\label{fig:abs-MAP}
\end{figure}

As seen in Fig.~\ref{fig:abs-MAP}, there is good agreement between BBN
calculations and astrophysical observations for the Deuterium and $^4$He
abundances~\cite{cfos}. The agreement for $^7$Li is less striking, though
not disastrously bad~\footnote{It seems unlikely that the low abundance of
$^7$Li observed could have been modified significantly by the decays of
heavy particles~\cite{EOV}: it would be valuable to refine the
astrophysical determinations.}. The good agreement between the
corresponding determinations of $\Omega_{Baryons}$ obtained from CMB and
Big-Bang nucleosynthesis calculations in conventional homogeneous
cosmology imposes important constraints on inhomogeneous models of
nucleosynthesis. In particular, they exclude the possibility that
$\Omega_{Baryons}$ might constitute a large fraction of $\Omega_{Tot}$.
Significant inhomogeneities might have been generated at the quark-hadron
phase transition, if it was strongly first-order~\cite{inhomog}. Although,
as already discussed, lattice calculations suggest that this is rather
unlikely, heavy-ion collision experiments must be the final arbiter on the
nature of the quark-hadron phase transition.

\section{Generating the Matter in the Universe}

As was pointed out by Sakharov~\cite{Sakharov}, there are three essential
requirements for generating the matter in the Universe via microphysics.
First, one needs a difference between matter and antimatter interactions,
as has been observed in the laboratory in the forms of violations of C and
CP in the weak interactions. Secondly, one needs interactions that violate
the baryon and lepton numbers, which are present as non-perturbative
electroweak interactions and in grand unified theories, but have not yet
been seen. Finally, one needs a breakdown of thermal equilibrium, which is
possible during a cosmological phase transition, for example at the GUT or
electroweak scale, or in the decays of heavy particles, such as a heavy
singlet neutrino $\nu_R$~\cite{leptog}. The issue then is whether we will
be able to calculate the resulting matter density in terms of laboratory
measurements.  Unfortunately, the Standard Model C and CP violation
measured in the quark sector seem unsuitable for baryogenesis, and the
electroweak phase transition in the Standard Model would have been second
order. However, additional CP violation and a first-order phase transition
in an extended electroweak Higgs sector might have been able to generate
the matter density~\cite{ewbaryog}, and could be testable at the LHC
and/or ILC. An alternative is CP violation in the lepton sector, which
could be probed in neutrino oscillation experiments, albeit indirectly, or
possibly in the charged-lepton sector, which might be related more
directly to the matter density~\cite{ER}.

In any case, detailed knowledge of the QGP equation of state would be
necessary if one were ever to hope to be able to calculate the
baryon-to-entropy ratio with an accuracy of a few percent.

\section{The Formation of Structures in the Universe}

The structures seen in the Universe - clusters, galaxies, stars and
eventually ourselves - are all thought to have developed from primordial
fluctuations in the CMB. This idea is supported visually by observations
of galaxies, which look smooth at the largest scales at high redshifts,
but cluster at smaller scales at low redshifts~\cite{2dF}. This scenario
requires amplification of the small fluctuations observed in the CMB,
which is possible with massive non-relativistic weakly-interacting
particles. On the other hand, relativistic light neutrinos would have
escaped from smaller structures, and so are disfavoured as amplifiers.
Non-relativistic `cold dark matter' is preferred, as seen in a comparison
of the available data on structures in the Universe with the cosmological
Concordance Model~\cite{LSS}.

\begin{figure}[htb]
\hspace*{-10mm}
\begin{center}
\includegraphics[height=.3\textheight, angle=0]{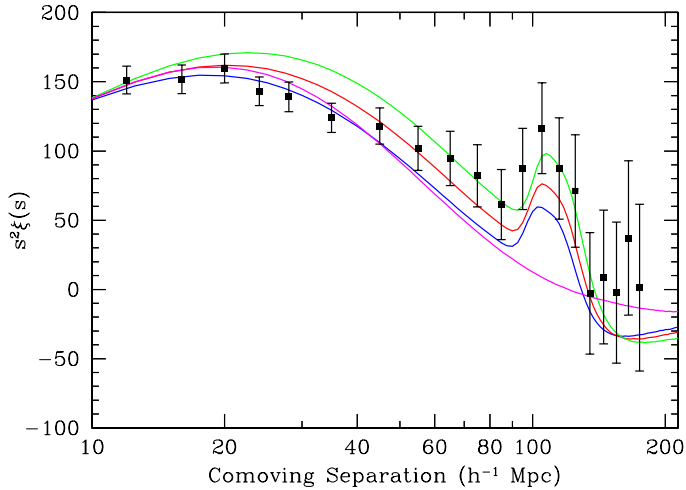}
\end{center}
\caption{\it
The baryonic `ripple' in the
large-scale correlation function of luminous red galaxies
observed in the Sloan Digital Sky Survey of galactic 
redshifts~\protect\cite{ripples}.}
\vspace*{0.5cm}
\label{fig:ripple}
\end{figure}

The hot news in the observational tests of this scenario has been the
recent detection of baryonic ripples from the Big Bang~\cite{ripples}, as
seen in Fig.~\ref{fig:ripple}. These are caused by sound waves spreading
out from irregularities in the CMB, which show up in the correlation
function between structures in the (near-)contemporary Universe as
features with a characteristic size. In addition to supporting the
scenario of structure formation by amplification of CMB fluctuations,
these observations provide measurements of the expansion history and
equation of state of the Universe.

\section{Do Neutrinos Matter?}

\begin{figure}[htb]
\hspace*{-10mm}
\centerline{\epsfxsize = 0.6\textwidth \epsffile{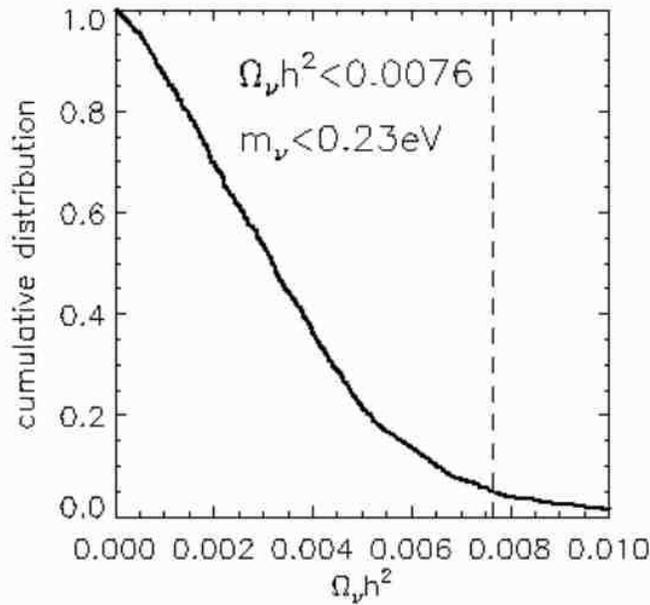}}
\caption{\it  
The likelihood function for the total neutrino density $\Omega_\nu h^2$ 
derived by WMAP~\protect\cite{Spergel}. The upper limit $m_\nu <
0.23$~eV applies if there are three degenerate neutrinos.}
\vspace*{0.5cm}
\label{fig:RS5}
\end{figure}

Oscillation experiments tell us that neutrinos have very small but
non-zero masses~\cite{atmo,solar}, and so must make up at least some of
the dark matter.  As already mentioned, since such light neutrinos move
relativistically during the epoch of structure formation, they would have
escaped from galaxies and not contributed to their formation, whereas they
could have contributed to the formation of clusters. Conversely, the
success of the cosmological Concordance Model enables one to set a
cosmological upper limit on the sum of light neutrino masses, as seen in
Fig.~\ref{fig:RS5}:  $\Sigma_\nu m_\nu < 0.7$~eV~\cite{Spergel}, which is
considerably more sensitive than direct laboratory searches. In the
future, this cosmological sensitivity might attain the range indicated by
atmospheric neutrino data~\cite{atmo}.

However, even if no dark matter effect of non-zero light neutrino masses
is observed, this does not mean that neutrinos have no cosmological role,
since unstable heavier neutrinos might have generated matter via the
Sakharov mechanism~\cite{Sakharov}.

\section{Particle Dark Matter Candidates}

Candidates for the non-relativistic cold dark matter required to amplify
CMB fluctuations include the {\it axion}~\cite{CAST}, TeV-scale {\it
weakly-interacting massive particles} (WIMPs) produced thermally in the
early Universe, such as the lightest supersymmetric partner of a Standard
Model particle (probably the lightest neutralino $\chi$), the {\it
gravitino} (which is likely mainly to have been produced in the very early
Universe, possibly thermally), and {\it superheavy relic particles} that
might have been produced non-thermally in the very early
Universe~\cite{Kolb} (such as the `cryptons' predicted in some string
models~\cite{cryptons}).

\section{Supersymmetric Dark Matter}

Supersymmetry is a very powerful symmetry relating fermionic `matter'
particles to bosonic `force' particles~\cite{susy}. Historically, the
original motivations for supersymmetry were purely theoretical: its
intrinsic beauty, its ability to tame infinities in perturbation theory,
etc. The first phenomenological motivation for supersymmetry at some
accessible energy was that it might also help explain the electroweak mass
scale, by stabilizing the hierarchy of mass scales in
physics~\cite{hierarchy}. It was later realized also that the lightest
supersymmetric particle (LSP) would be stable in many models~\cite{Fayet}.
Moreover, it should weigh below about 1000 GeV, in order to stabilize the
mass hierarchy, in which case its relic density would be similar to that
required for cold dark matter~\cite{EHNOS}. As described below,
considerable effort is now put into direct laboratory searches for
supersymmetry, as well as both direct and indirect astrophysical searches.

Here I concentrate on the minimal supersymmetric extension of the Standard
Model (MSSM), in which the Standard Model particles acquire superpartners
and there are two doublets of Higgs fields. The interactions in the MSSM
are completely determined by supersymmetry, but one must postulate a
number of soft supersymmetry-breaking parameters, in order to accommodate
the mass differences between conventional particles and their
superpartners. These parameters include scalar masses $m_0$, gaugino
masses $m_{1/2}$, and trilinear soft couplings $A_0$. It is often assumed
that these parameters are universal, so that there is a single $m_0$, a
single $m_{1/2}$, and a single $A_0$ parameter at the input GUT scale, a
scenario called the constrained MSSM (CMSSM). However, there is no deep
theoretical justification for this universality assumption, except in
minimal supergravity models. These models also make a prediction for the
gravitino mass: $m_{3/2} = m_0$, which is not necessarily the case in the
general CMSSM.

As already mentioned, the lightest supersymmetric particle is stable in
many models, this because of the multiplicative conservation of $R$
parity, which is a combination of spin $S$, lepton number $L$ and baryon
number $B$: $R = (-1)^{2S - L + 3B}$.  It is easy to check that
conventional particles have $R = +1$ and sparticles have $R = -1$. As a
result, sparticles are always produced in pairs, heavier sparticles decay
into lighter ones, and the lightest supersymmetric particle (LSP) is
stable.

The LSP cannot have strong or electromagnetic interactions, because these
would bind it to conventional matter, creating bound states that would be
detectable as anomalous heavy nuclei. Among the possible
weakly-interacting scandidates for the LSP, one finds the {\it sneutrino},
which has been excluded by a combination of LEP and direct searches for
astrophysical dark matter, the lightest {\it neutralino} $\chi$, and the
{\it gravitino}. There are good prospects for detecting the neutralino or
gravitino in collider experiments, and neutralino dark matter may also be
detectable either directly or indirectly, but gravitino dark matter would
be a nightmare for detection.

\section{Constraints on Supersymmetry}

\begin{figure}[h]
\includegraphics[height=3in]{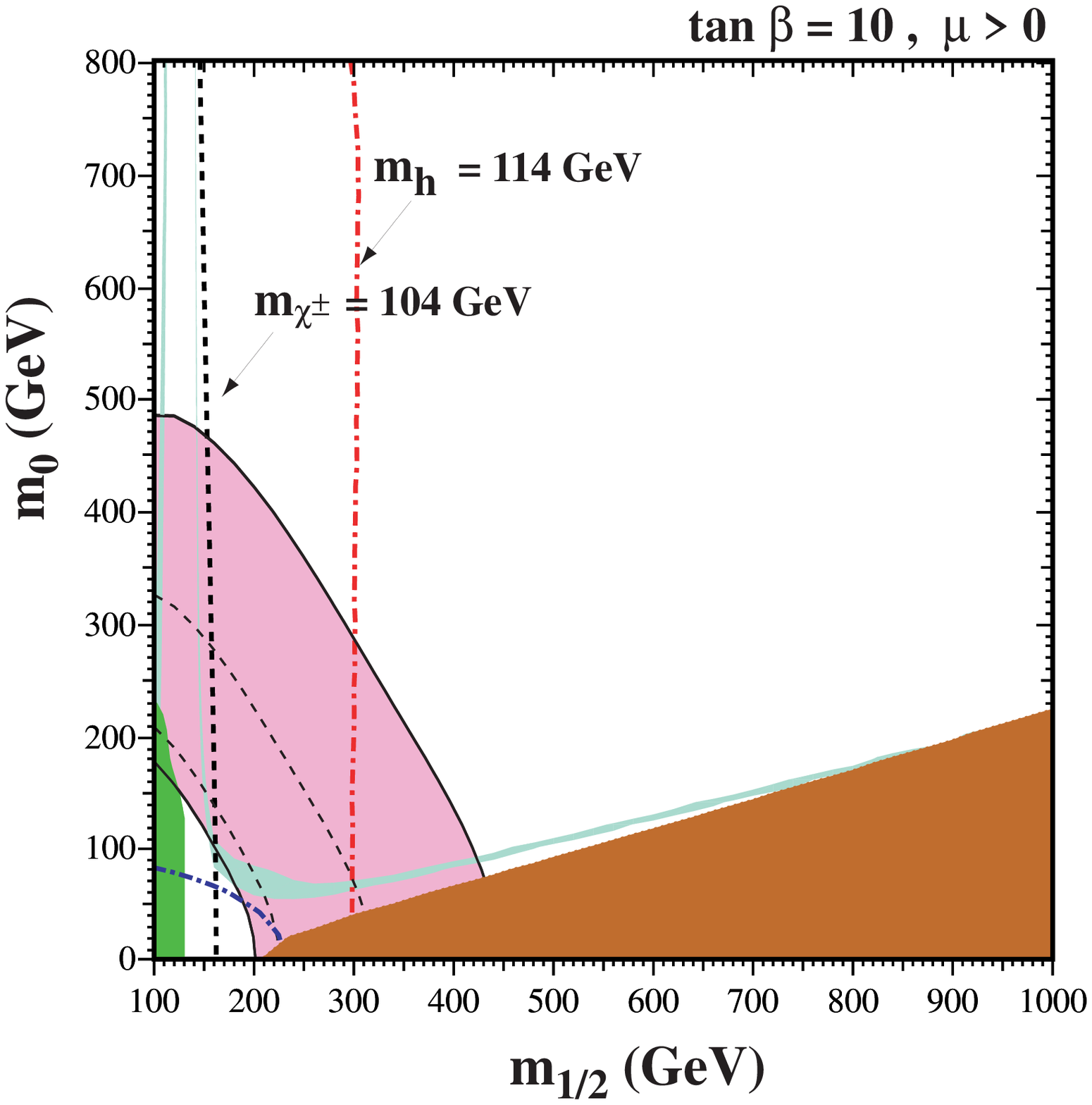}
\includegraphics[height=3in]{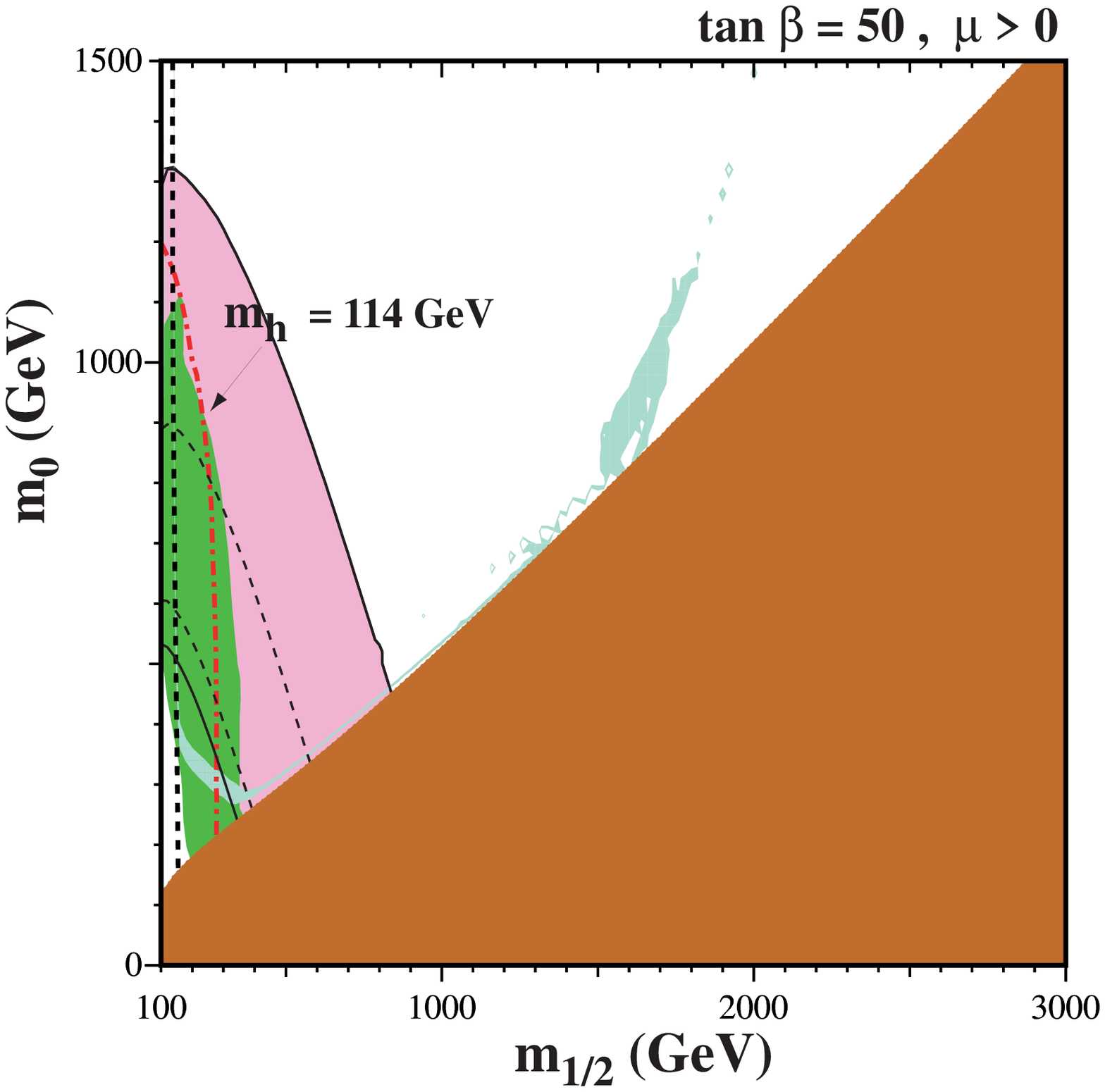}
\caption{\label{fig:CMSSMDM}
{\it The $(m_{1/2}, m_0)$ planes for  (a) $\tan \beta = 10$ and $\tan 
\beta = 50$ with $\mu > 
0$ and $A_0 = 0, assuming m_t = 175$~GeV and
$m_b(m_b)^{\overline {MS}}_{SM} = 4.25$~GeV. The near-vertical (red) 
dot-dashed lines are the contours $m_h = 114$~GeV, and the near-vertical 
(black) dashed
line is the contour $m_{\chi^\pm} = 104$~GeV. Also
shown by the dot-dashed curve in the lower left is the corner
excluded by the LEP bound of $m_{\tilde e} > 99$ GeV. The medium (dark
green) shaded region is excluded by $b \to s
\gamma$, and the light (turquoise) shaded area is the cosmologically
preferred regions with \protect\mbox{$0.094 \le \Omega_\chi h^2 \le 
0.129$}. In 
the dark  
(brick red) shaded region, the LSP is the charged ${\tilde \tau}_1$. The
region allowed by the E821 measurement of $a_\mu$ at the 2-$\sigma$  
level, is shaded (pink) and bounded by solid black lines, with dashed
lines indicating the 1-$\sigma$ ranges~\protect\cite{EOSS}.}}
\end{figure}

Important constraints on supersymmetry are imposed by the absences of
sparticles at LEP and the Tevatron collider, implying that sleptons and
charginos should weigh $> 100$ GeV~\cite{LEPsusy}, and that squarks and
gluinos should weigh $> 250$ GeV, respectively. Important
indirect constraints are imposed by the LEP lower limit on the mass of the
lightest Higgs boson, 114 GeV~\cite{LEPHiggs}, and the experimental
measurement of $b \to s \gamma$ decay~\cite{bsg}, which agrees with the
Standard Model. The measurement of the anomalous magnetic moment of the
muon, $g_\mu - 2$, also has the potential to constrain supersymmetry, but
the significance of this constraint is uncertain, in the absence of
agreement between the $e^+ e^-$ annihilation and $\tau$ decay data used to
estimate the Standard Model contribution to $g_\mu - 2$~\cite{g-2}.
Finally, one of the strongest constraints on the supersymmetric parameter
space is that imposed by the density of dark matter inferred from
astrophysical and cosmological observations. If this is composed of the
lightest neutralino $\chi$, one has $0.094 < \Omega_\chi h^2 <
0.129$~\cite{Spergel}, and it cannot in any case be higher than this. For
generic domains of the supersymmetric parameter space, this range
constrains $m_0$ with an accuracy of a few per cent as a function of
$m_{1/2}$, as seen in Fig.~\ref{fig:CMSSMDM}~\cite{EOSS}.

\section{The Relic Density and the Quark-Gluon Plasma}

\begin{figure}
\begin{center}
\includegraphics[width=0.6\hsize]{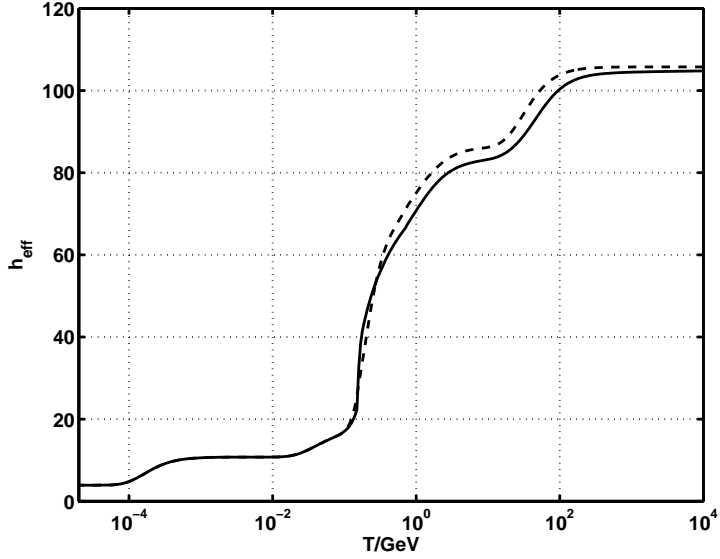}
  \caption{\label{f:dofs}
The factor $h_{\rm eff}(T)$
calculated using different equations of state~\protect\cite{Hind}.}
\end{center}  
\end{figure}

The accurate calculation of the relic density depends not only on the
supersymmetric model parameters, but also on the effective Hubble
expansion rate as the relic particles annihilate and freeze out of thermal
equilibrium~\cite{EHNOS}:
$$
\dot{n} + 3 Hn = - <\sigma_{ann} v > \, (n^2 - n^2_{eq}).
$$
This is, in turn, sensitive to the effective number of particle species:
$$
Y_0 \simeq \left ( \frac{45}{\pi} \right )^{\frac{1}{2}} \; 
\frac{1}{m M_P <\sigma_{ann} v >_{T_f} g_*^{1/2} (T_f)}
$$
where
$$
g_*^{1/2} (T) = 
\frac{H_{eff}}{g_{eff}}^{\frac{1}{2}} \left ( 1 + \frac{T}{3} \frac{d 
{\rm ln} h_{eff}}{dT} \right ).
$$
To calculate this at the per-cent level, one needs to understand with
comparable accuracy the equation of state of hot QCD around the 
freeze-out temperature~\cite{Hind}, which is typically $\sim m_\chi / 20 
\sim$ a few 
GeV. Fig.~\ref{f:dofs} the comparison between the predictions of different 
equations of state for $h_{eff}$: it would be good to reach consensus on 
at the per-cent level.

\section{Supersymmetry Searches at the LHC}

\begin{figure}
\begin{center}
\mbox{\epsfig{file=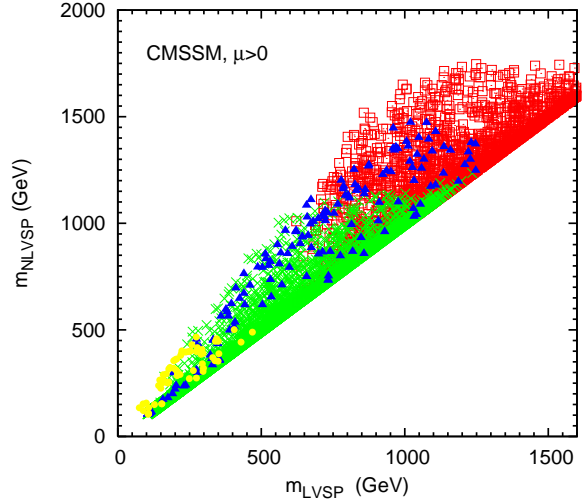,height=6.8cm}}
\end{center}
\caption{\it
Scatter plot of the masses of the lightest visible
supersymmetric particle (LVSP) and the next-to-lightest visible
supersymmetric particle (NLVSP) in the CMSSM. The darker (blue)
triangles satisfy all the laboratory, astrophysical and cosmological
constraints. For comparison, the dark (red) squares and medium-shaded
(green) crosses respect the laboratory
constraints, but not those imposed by astrophysics and cosmology.
In addition, the (green) crosses represent models which are expected to be
visible at the LHC. The
very light (yellow) points are those for which direct detection of
supersymmetric dark matter might be possible~\protect\cite{reach}.}
\label{fig:LHC}
\end{figure}

A `typical' CMSSM event should be relatively `easy' to see at the LHC, as
it should have a lot of `missing' transverse energy carried away by
invisible neutralinos, accompanied by several jets and/or leptons. Many
studies have shown that the LHC should be able to cover most of the region
of parameter space allowed by cosmology~\cite{Bench1}, as seen in
Fig.~\ref{fig:LHC}~\cite{reach}.  Moreover, the LHC should be able to see
several species of supersymmetric particles, which might make possible the
{\it a priori} calculation of the relic density with an uncertainty
comparable to that provided by astrophysical and cosmological
data~\cite{Bench2}. Minimal supergravity models might be even easier to
discover, since in generic domains of parameter space the next-to-lightest
supersymmetric particle would live a long time before it decayed into the
gravitino~\cite{GDM}, liberating a large amount of energy that could be
measured in a dedicated detector~\cite{Feng}.

\section{Strategies for Detecting Supersymmetric Dark Matter}

These include searches for the annihilations of relic particles in the
galactic halo:$\chi \chi \to$ antiprotons or positrons, annihilations in
the galactic centre: $\chi \chi \to \gamma + \dots $, annihilations in the
core of the Sun or the Earth: $\chi \chi \to \nu + \cdots \to \mu +
\cdots$ , and scattering on nuclei in the laboratory: $\chi A \to chi A$.

\begin{figure}[htb]
\centerline{\epsfxsize = 0.5\textwidth \epsffile{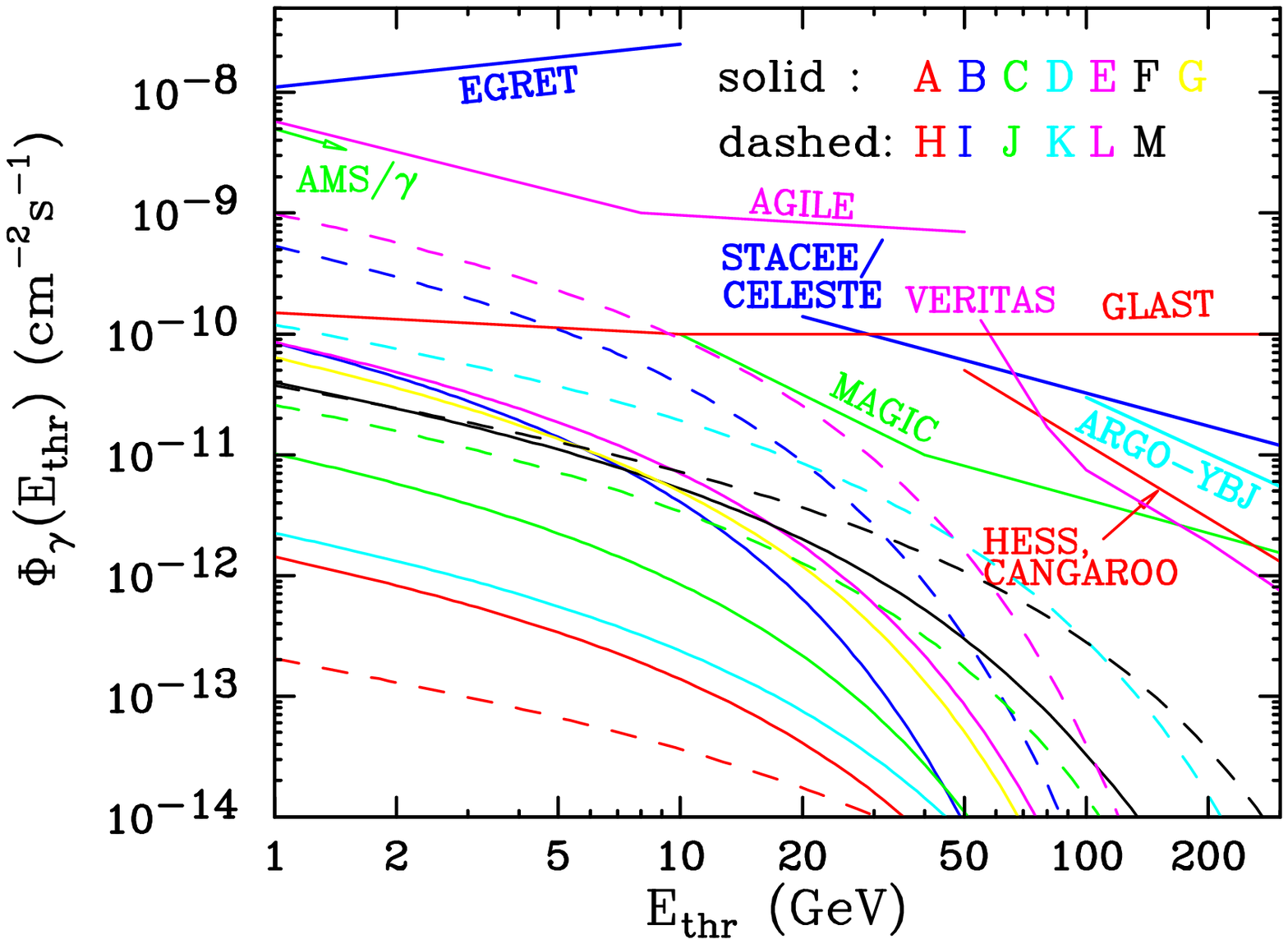} 
\hfill \epsfxsize = 0.5\textwidth \epsffile{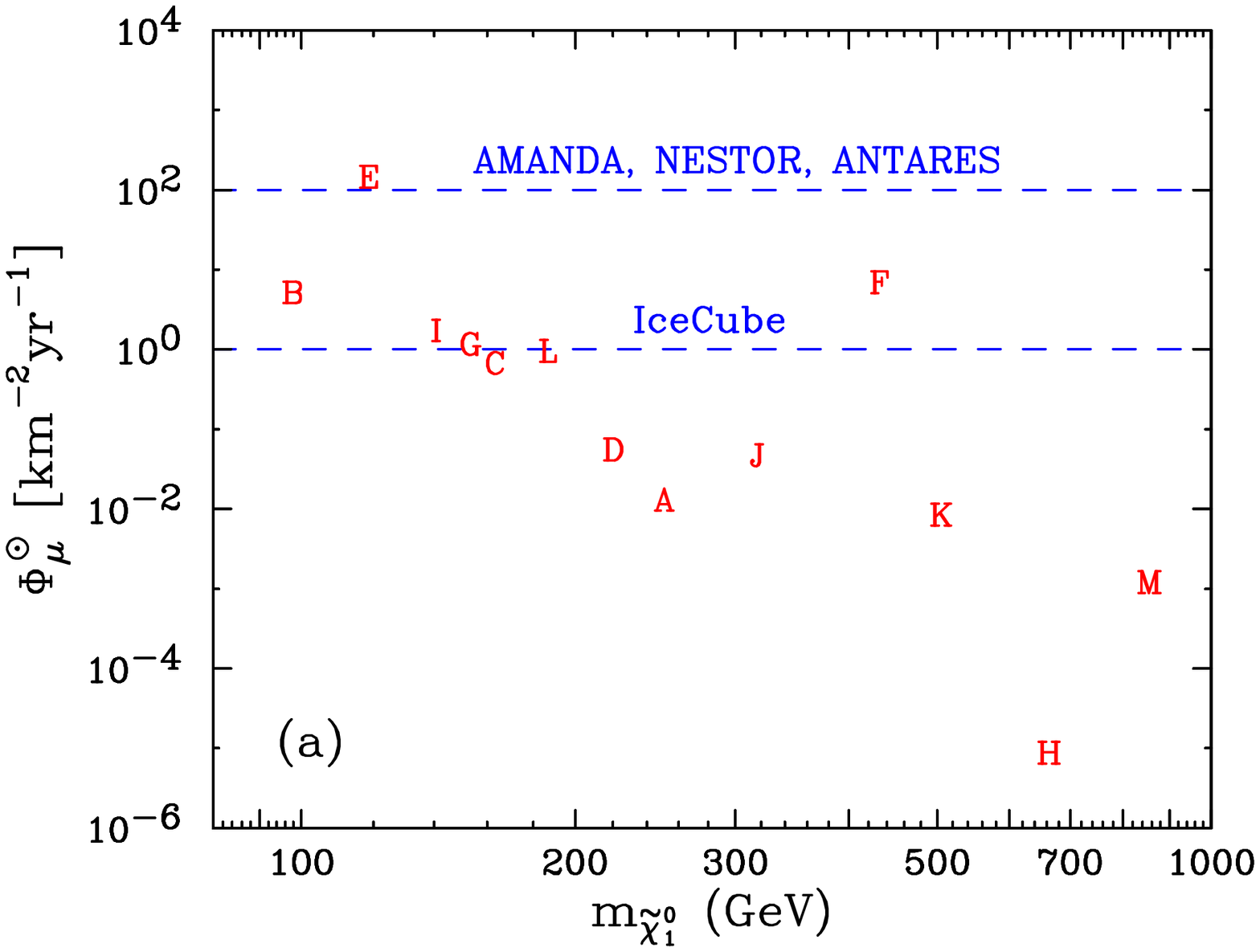} 
}
\caption{\it
Left panel: Spectra of photons from the annihilations of
dark matter particles in the core of our galaxy, in different benchmark
supersymmetric models~\protect\cite{EFFMO}. Right panel: Signals for
muons produced by energetic neutrinos originating from annihilations of 
dark matter particles in the core of the Sun, in the same benchmark
supersymmetric models~\protect\cite{EFFMO}.}
\vspace*{0.5cm}
\label{fig:RS13}
\end{figure}

After some initial excitement, recent observations of cosmic-ray
antiprotons are consistent with production by primary matter cosmic rays.
Moreover, the spectra of annihilation positrons calculated in a number of
CMSSM benchmark models~\cite{Bench1} seem to fall considerably below the
cosmic-ray background~\cite{EFFMO}. Some of the spectra of photons from
annihilations in Galactic Centre, as calculated in the same set of CMSSM
benchmark scenarios, may rise above the expected cosmic-ray background,
albeit with considerable uncertainties due to the unknown enhancement of
the cold dark matter density. In particular, the GLAST experiment may have
the best chance of detecting energetic annihilation photons~\cite{EFFMO},
as seen in the left panel of Fig.~\ref{fig:RS13}. Annihilations in the
Solar System also offer detection prospects in some of the benchmark
scenarios, particularly annihilations inside the Sun, which might be
detectable in experiments such as AMANDA, NESTOR, ANTARES and particularly
IceCUBE, as seen in the right panel of Fig.~\ref{fig:RS13}~\cite{EFFMO}.

\begin{figure}
\begin{center}
\mbox{\epsfig{file=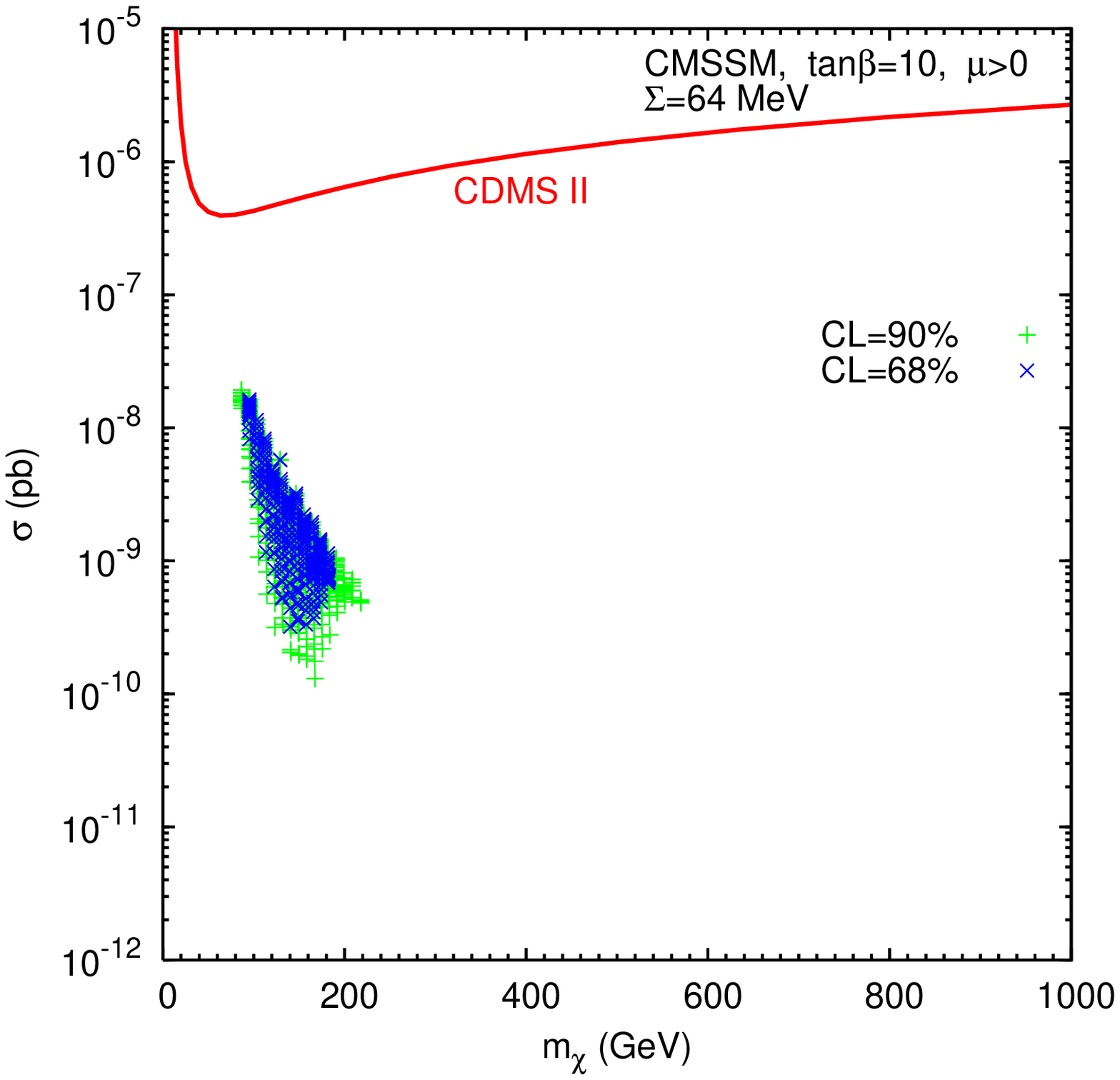,height=7cm}}
\mbox{\epsfig{file=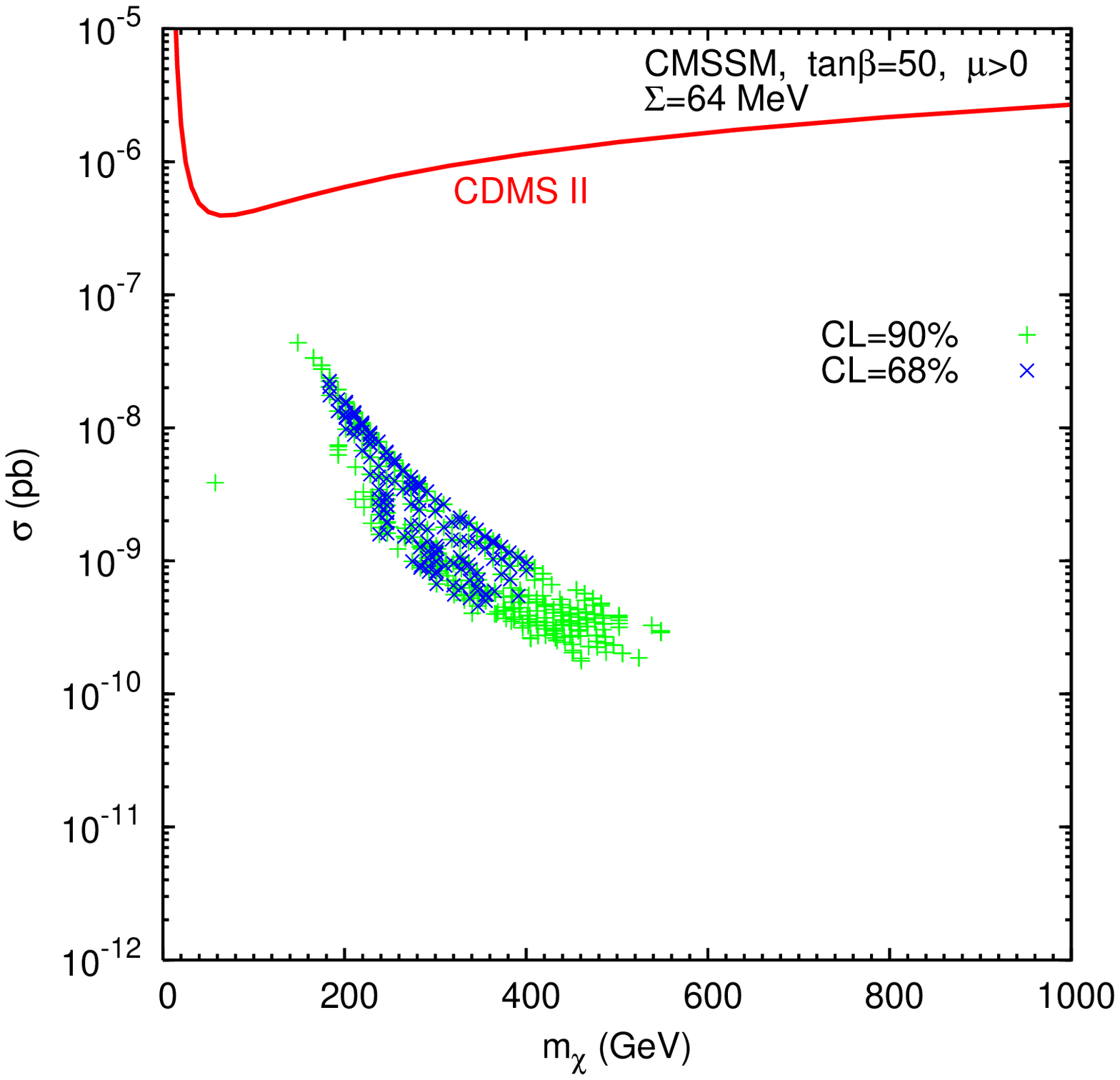,height=7cm}}
\end{center}
\caption{\label{fig:EHOW3}
{\it Scatter plots of the spin-independent elastic-scattering cross
section predicted in the CMSSM for (a) $\tan \beta = 10, \mu > 0$ and
(b) $\tan \beta = 50, \mu > 0$, each with $\sigma_{\pi N} =
64$~MeV~\protect\cite{EOSS5}. The predictions for models allowed at the 
68\% (90\%) confidence
levels~\cite{EHOW3} are shown by blue $\times$ signs (green $+$ signs).}}
\end{figure}

The rates for elastic dark matter scattering cross sections calculated in
the CMSSM are typically considerably below the present upper limit imposed
by the CDMS II experiment, in both the benchmark scenarios and the global
fit to CMSSM parameters based on present data~\cite{EOSS5}. However, if
the next generation of direct searches for elastic scattering can reach a
sensitivity of $10^{-10}$ pb, they should be able to detect supersymmetric
dark matter in many supersymmetric scenarios. Fig.~\ref{fig:EHOW3}
compares the cross sections calculated under a relatively optimistic
assumption for the relevant hadronic matrix element $\sigma_{\pi N} =
64$~MeV, for choices of CMSSM parameters favoured at the 68~\% (90~\%)  
confidence level in a recent analysis using the observables $m_W, \sin^2
\theta_W, b \to s \gamma$ and $g_\mu - 2$~\cite{EHOW3}.

\section{Connections between the Big Bang and Little Bangs}

Astrophysical and cosmological observations during the past few years have
established a Concordance Model of cosmology, whose matter content is
quite accurately determined. Most of the present energy density of the
Universe is in the form of dark vacuum energy, with about 25 $\%$ in the
form of dark matter, and only a few $\%$ in the form of conventional
baryonic matter. Two of the most basic questions raised by this
Concordance Model are the nature of the dark matter and the origin of
matter.

Only experiments at particle colliders are likely to be able to answer
these and other fundamental questions about the early Universe. In
particular, experiments at the LHC will recreate quark-gluon plasma
conditions similar to those when the Universe was less than a microsecond
old~\cite{ALICE}, and will offer the best prospects for discovering
whether the dark matter is composed of supersymmetric
particles~\cite{ATLAS,CMS}. LHC experiments will also cast new light on
the cosmological matter-antimattter asymmetry~\cite{LHCb}. Moreover,
discovery of the Higgs boson will take us closer to the possibilities for
inflation and dark energy.

There are many connections between the Big Bang and the little bangs we
create with particle colliders. These connections enable us both to learn
particle physics from the Universe, and to use particle physics to
understand the Universe.

\end{document}